\documentclass[a4paper,12pt]{article}
\usepackage[utf8]{inputenc}
\usepackage[english]{babel}
\usepackage{authblk}
\usepackage{graphicx}
\usepackage{mathptmx}
\usepackage[singlespacing]{setspace}
\usepackage[headheight=1in,margin=1in]{geometry}
\usepackage{fancyhdr}
\usepackage{lipsum}
\usepackage{array}
\usepackage{booktabs}

\makeatletter
\def\@maketitle{%
  \newpage

  \begin{center}%
  \let \footnote \thanks
    {\LARGE \@title \par}%
    \vskip 1em
    {\large \@author \par}%
    \vskip 0.5em
    {\normalsize Pew Research Center, Washington, DC, USA \par}%
  \end{center}%
  \par
  \vskip 0.1em}
\makeatother

\graphicspath{{images/}}

\title{Investigating Click Behaviors On Google Search \\Result Pages That Produce an AI Overview}
\author{Athena Chapekis, Anna Lieb, Sono Shah, Aaron Smith}

\date{}

\begin{document}

\maketitle
\thispagestyle{fancy}

\begin{center}
\end{center}

\section*{Extended Abstract}

In 2024, Google introduced ``AI Overviews," a feature that displays an AI-generated result summary at the top of many Google search pages. This study investigates the role of AI in Google search using one month of web browsing data from a representative panel of 900 U.S. adults. Our analysis of the panelists' Google searches sheds light on AI Overviews, when they appear in Google search results, and what user behaviors are associated with AI Overviews. We identify several attributes that make a search query more likely to generate an AI Overview, including the length of a query, whether the query begins with a question word, and whether the query contains both a noun and verb. When it comes to user behavior, we find that clicks to sources cited in AI Overviews are very rare, occurring in only about 1\% of visits to AI Overviews. We also find that AI Overviews are associated with fewer clicks and higher rates of ending browsing sessions. Importantly, results from a mixed-effects logistic regression model indicate that these associations hold when controlling for random effects by panelist and query attributes that make AI Overviews more likely to appear. 

\textbf{Related Work.} Generative AI in search is a relatively recent innovation, but search engine result page (SERP) features that surface summaries, snippets, and other algorithmically curated content alongside result links have been a growing part of Google Search for over a decade \cite{sullivan_reintroduction_2018, oliveira_evolution_2023}. Previous research has found that these search components can decrease click-through rates, influence users' credibility assessments, and capture user attention \cite{chilton_addressing_2011, lurie_investigating_2018, marcos_effect_2015, sushmita_factors_2010}. In the past year, some online publishers have attributed declining web traffic to AI summaries replacing traditional search results \cite{simonetti_news_2025}. Since Google AI Overviews are a relatively new search feature, there are many opportunities for research in this area. Recently, Hu et al. (2025) found that AI Overviews appeared in 84\% of baby care and pregnancy search results. Further, they found that these AI Overviews gave inconsistent information and often lacked medical safeguards or warnings \cite{hu_auditing_2025}. Other researchers have expressed concerns that generative search could contribute to biased information seeking behaviors, less critical evaluation of sources, and conditions where users are increasingly reliant on a small set of sources \cite{sharma_generative_2024, venkit_search_2025}. Our study's approach using participants' web browsing data is most similar to past research conducted by Gleason et al (2023), although we focus on AI Overviews rather than featured snippets and other extracted results \cite{gleason_google_2023}. Our study also includes an analysis of search query attributes, which is an essential inclusion as previous work has shown that attributes such as query intent, query length, and parts of speech are important factors in user search behaviors \cite{hafernik_understanding_2013, bendersky_analysis_2009}.

\begin{figure*}[ht]
\centering
\includegraphics[width=0.99\textwidth]{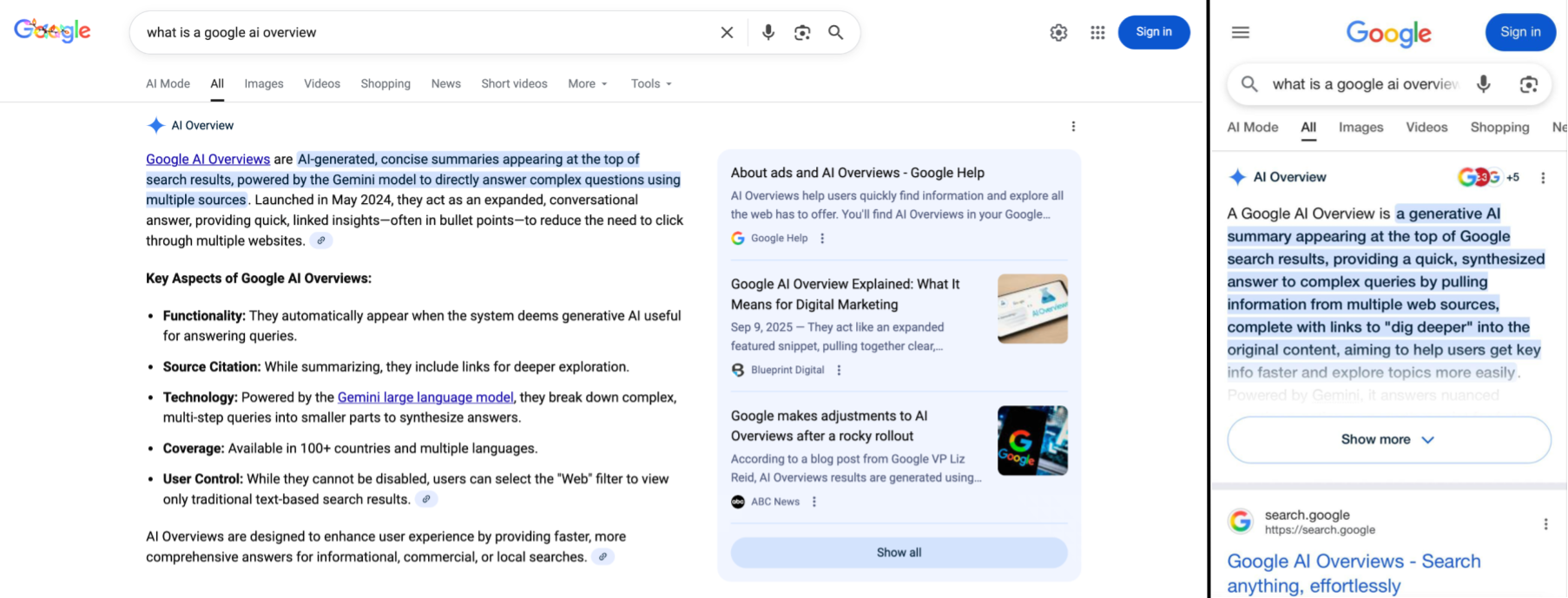} 
\caption{In a desktop web browser (left), Google AI Overviews highlight up to three sources on the right with a ``Show all'' button to list more. The number of sources listed depends on screen size and result format. For example, a mobile device (right) may not display any sources until the user clicks ``Show more."}
\label{sample-overview}
\end{figure*}

\textbf{Methods.} To investigate how users interact with AI search summaries, we analyzed one month of web browsing data from a sample of 900 U.S. adults. The sample includes members of the Ipsos KnowledgePanel Digital online panel, who were recruited with an address-based sampling methodology to ensure that the panel was as representative as possible \cite{harter_aapor_2016}. Those who consented to the study and installed the required software on their personal devices were monitored from March 1 to 31, 2025. Over the one-month period, the 900 panelists conducted 2,457,176 web page visits; 91,121 of those visits were to Google Search pages, amounting to 68,879 distinct Google search queries. On April 10-11 2025, we collected SERPs for all Google Search URLs that appeared in the dataset. For SERPs that include an AI Overview, we collected the AI Overview text and up to three sources cited in the AI Overview. As of April 2025, three sources was the maximum number of sources that was highlighted on a desktop web browser SERP without clicking to see more (see Figure \ref{sample-overview} for an example). 

Next, we used panelist activity URLs, corresponding timestamps, and the collected SERPs to identify panelists' next action after visiting each Google Search page. The user action categories were defined as follows: (1) ``Clicking on a link in the AI Overview" if the next-visited URL matched any of the first three links cited in a page’s AI Overview; (2) ``Clicking on a link from the search results" if the next-visited URL matched any links displayed on the first page of search results; (3) ``Continuing to search Google" if the next-visited URL was a different Google search page; (4) ``Leaving Google to browse a different site" if the next-visited URL did not appear in the search results or AI Overview sources; and (5) ``Ending their browsing session" if the user exited the web browser for five seconds or more.

Since this study uses observational rather than experimental data, we are unable to make causal claims about the relationship between AI Overviews and user behaviors. Despite these limitations, our observational data can be used to analyze user behaviors associated with search results with and without AI Overviews. We used a mixed effects model to test whether the differences in user behavior still hold when controlling for key query attributes and random effects of user-specific tendencies. We fit two models to compare user behaviors after visiting pages with and without AI Overviews. The first model predicts the likelihood of users clicking on results, and the second model predicts the likelihood of users ending the browsing session. Since we wanted to incorporate both fixed and random effects into our model with a binary outcome variable, we used the \texttt{glmer} function as implemented in the \texttt{lme4} R package to fit a logistic mixed-effects model \cite{bates_lme4_2025}. The covariates for our models are shown in Table \ref{glmm-results}, with fixed effects for query attributes, random effects by panelist, and the binary outcome as either clicking a result (Model 1) or ending the browsing session (Model 2).

\begin{figure}[h]
\centering
\includegraphics[width=0.7\textwidth]{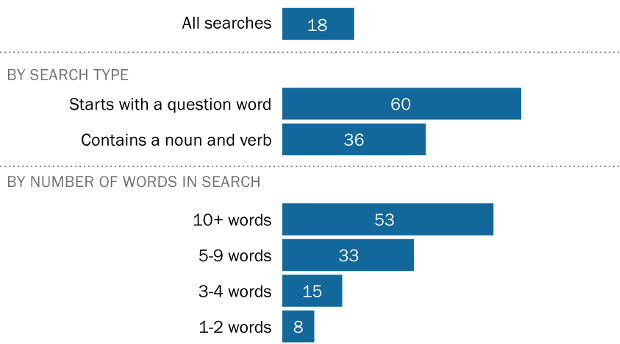} 
\caption{Percent of Google searches that produced an AI Overview in March 2025. Google search queries that contain more words, ask questions, or use full sentences tend to produce AI summaries more often (p \textless 0.05).}
\label{query-attributes}
\end{figure}

\begin{figure}[ht]
\centering
\includegraphics[width=0.7\columnwidth]{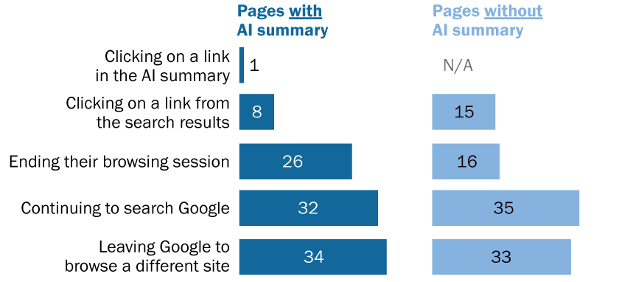} 
\caption{Percent of Google searches that resulted in each user action. When users encounter AI Overviews, they are significantly less likely to click on a link from the results and more likely to end their browsing session (p \textless 0.01).}
\label{next-action-chart}
\end{figure}

\textbf{Results.} Overall, we find that about 18\% of all Google searches in our study generated an AI Overview. We observe that longer queries, queries that begin with a question word, and queries that contain both a noun and a verb tend to produce an AI Overview in the results (see Figure \ref{query-attributes}). When it comes to user click behavior, our panel data indicate that users very rarely click on links to sources cited within AI Overviews. Just 1\% of all visits to pages with an AI Overview resulted in a click to the cited sources. Additionally, we find that Google users who encounter an AI Overview are less likely to click on links to other websites than users who do not see one. Users clicked on a search result link in 15\% of all visits to SERPs without AI Overviews. But on pages that generated AI Overviews, users clicked on results nearly half as often (8\% of visits). We also find that Google users are more likely to end their browsing session after visiting a search page with an AI Overview compared to pages without one. Users in our study ended their browsing session on 26\% of pages with an AI Overview, compared to 16\% of pages without an Overview (see Figure \ref{next-action-chart}). Based on our regression analysis, differences in click-through rates and ending the session remained significant, even when controlling for key query attributes (see Table \ref{glmm-results}). The patterns that we observe about AI Overviews, when they appear, and how users interact with them have important implications for how Google users seek out information and what sources they encounter online.

\bibliographystyle{plain}
\bibliography{sources.bib}

\newpage

\section*{Figures and Tables}

\begin{table}[ht]
    \centering
    \setlength{\extrarowheight}{4pt}
    \begin{tabular}{>{\centering\arraybackslash}p{0.3\linewidth}>{\centering\arraybackslash}p{0.15\linewidth}>{\centering\arraybackslash}p{0.15\linewidth}}\toprule
         &  Model 1: 
Click result& Model 2: 
End session\\\midrule
         Intercept&  -2.7213***
(0.0773)& -1.8255***
(0.0375)\\
         AI Overview present&  -0.6805***
(0.0510)& 0.4743***
(0.0356)\\
         Question query&  -0.1518**
(0.0689)& 0.3454***
(0.0535)\\
         Query has noun and verb&  0.0372
(0.0293)& 0.0596**
(0.0273)\\
         Query length (words)&  -0.0095***
(0.0037)& 0.0025
(0.0033)\\
         AI Overview present $*$ Question query&  0.1226
(0.1040)& -0.3089***
(0.0716)\\
         AI Overview present $*$ Query has noun and verb&  0.1240*
(0.0734)& 0.1051**
(0.0521)\\
         AI Overview present $*$ Query length&  0.0194***
(0.0063)& -0.0012
(0.0055)\\
         Random effect by panelist &  SD = 1.614& SD = 0.718\\
 N& 
91,121 obs,
647 panelists&91,121 obs,
647 panelists\\ \bottomrule
    \end{tabular}
    \caption{Logistic mixed effects models for user likelihood of clicking on a search result link (Model 1) and ending browsing session (Model 2). Estimates are in log-odds. (*), (**), and (***), denote significance at (p \textless 0.1), (p\textless 0.05), and (p \textless 0.01) respectively. }
    \label{glmm-results}
\end{table}

\end{document}